\documentclass[reprint,amsmath,amssymb,aps,prl]{revtex4-2}

\usepackage{graphicx}
\usepackage{dcolumn}%
\usepackage{bm}
\usepackage[colorlinks=true,linkcolor=blue,citecolor=blue,urlcolor=blue]{hyperref}
\usepackage[dvipsnames]{xcolor}

\begin{document}

\title{Isometric Incompatibility in Growing Elastic Sheets}

\author{Yafei Zhang}

\author{Michael Moshe}%
\email{michael.moshe@mail.huji.ac.il}

\author{Eran Sharon}
\email{erans@mail.huji.ac.il}

\affiliation{Racah Institute of Physics, The Hebrew University of Jerusalem, Jerusalem, 9190401, Israel.}%

\begin{abstract}

Geometric incompatibility, the inability of a material’s rest state to be realized in Euclidean space, underlies shape formation in natural and synthetic thin sheets. Classical Gauss and Mainardi-Codazzi-Peterson incompatibilities explain many patterns in nature, but they do not exhaust the mechanisms that frustrate thin elastic sheets. We identify a new incompatibility that forbids smooth stretching-free configurations, even when the rest state of the elastic sheet locally satisfies the Gauss and Mainardi-Codazzi-Peterson compatibility conditions. We demonstrate this principle in a model of surface growth with positive Gaussian curvature, where a geometric horizon forms, leading to the onset of frustration. Experiments, simulations, and theory show that the sheet responds by nucleating periodic d-cone-like dimples. We show that this obstruction to stretching-free configurations is topological, and we point to open questions concerning the origin of frustration.
\end{abstract}

\maketitle

\textit{Introduction—}Geometric frustration in thin sheets is ubiquitous across nature and synthetic systems, governing morphogenesis in living tissues~\cite{armon2011geometry,zhang2025geometrically,katifori2010foldable,dervaux2008morphogenesis,arroyo2014shape,latorre2018active}, defect organization in crystalline and amorphous solids~\cite{irvine2010pleats,moshe2015geometry,meng2014elastic}, and mechanical response of architectured and adaptive materials~\cite{wang2025geometrically,serafin2021frustrated,levin2020self}.  A characteristic expression of frustration is the generation of residual stresses, which in turn initiate instabilities resulting with intricate patterns~\cite{klein2007shaping,davidovitch2019geometrically,sun2021fractional,tobasco2022exact,timounay2020crumples,bartolo2019topological,hure2011wrapping,moguel2025topological}.  Over the past two decades, two  sources of geometric frustration have been identified for thin sheets: the Gauss frustration and the Mainardi-Codazzi-Peterson (MCP) frustration~\cite{efrati2009elastic,siefert2021euclidean}.  Both originate from the Gauss or MCP compatibility conditions, which must be satisfied by the metric and curvature tensors $a$ and $b$ of any actual configuration, but may be violated by the reference fields $\bar{a}$ and $\bar{b}$ defining the target geometry~\footnote{The actual metric and curvature tensors $a$ and $b$ encode the length of the line element and variation of the surface normal in the realized configuration.  In contrast, the reference tensors $\bar{a}$ and $\bar{b}$ specify the preferred length of the line element and variation of the surface normal in a stress-free geometry.}. In cases where $\bar{a}$ and $\bar{b}$ are \textit{geometrically incompatible}, every material element is intrinsically frustrated: it cannot be embedded in Euclidean space without residual stress, as follows from the fundamental theorem of surfaces \cite{do2016differential}.

An important property of geometrically incompatible $\bar{a}$ and $\bar{b}$ is that at equilibrium the actual forms $a$ and $b$ deviate from their rest values. For non-Euclidean plates, i.e., when $\bar{b}=0$ and $\bar{a}$ is non-Euclidean, it was rigorously proven that if and only if there exists an isometric embedding with finite bending energy \footnote{a $W^{2,2}$ embedding~\cite{lewicka2011scaling}}, then the limit of vanishing thickness $t$ drives the sheet toward a configuration that minimizes bending energy among all isometric embeddings with $a = \bar{a}$ \cite{efrati2009elastic,lewicka2011scaling}, and the elastic energy scales as $t^3$ (bending energy).  This mechanism underlies shape selection and shape transitions in a wide range of soft matter systems \cite{klein2007shaping,witten2007stress,armon2011geometry,siefert2021euclidean,zhang2025geometrically}.

Another origin of frustration is topological: closed sheets, such as vesicles, fruits and pollen grains, can become frustrated due to topological constraints even if they are locally compatible ~\cite{katifori2010foldable,jackson2023scaling,vellutini2025patterned,he2014apical,wang2023curvature}. For such surfaces, Gauss-Bonnet theorem determines a rigid connection between the topology of the surface and the integral of the Gaussian curvature over it~\cite{do2016differential}. Such topological constraints do not exist in open surfaces.
In this Letter, we identify a new form of incompatibility in open growing surfaces. We show that integrating more than a $4\pi$ \emph{positive} Gaussian curvature on a circular disk or an annulus leads to geometric frustration.
We show that regular configurations satisfying $a = \bar{a}$ with integrated curvatures less than $4\pi$ cannot be extended beyond this limit, irrespective of $\bar{b}$, even if $\bar{a}$ and $\bar{b}$ are locally Gauss- and MCP-compatible.

The inability to extend the isometric configurations, which violates the smooth embedding hypothesis in \cite{lewicka2011scaling}, introduces a fundamentally new mechanism of shape selection: it forces the sheet into complex configurations with diverging energy density as the thickness decreases. Similarly to Gauss and MCP incompatibilities, this new incompatibility has the potential to account for an entirely new class of pattern-forming phenomena in natural and synthetic slender solids.

To clarify the context of our results, it is important to note that a situation where $a$ significantly deviates from $\bar{a}$ in a thin sheet, regardless of $\bar{b}$, is not unprecedented. The canonical example arises  in the growth of a surface with constant negative curvature with the topology of a cylinder. It is well-known that this geometry possesses a horizon beyond which the symmetry breaks, and the emerging periodic pattern becomes increasingly refined as the thickness decreases \cite{sharon2010mechanics}.
This behavior has been observed in synthetic sheets and invoked to explain morphogenetic patterns in natural tissues~\cite{sharon2010mechanics,amar2012petal,marder2006geometry,arroyo2014shape}.
Importantly, this geometrically frustrated state does not reflect Gauss or MCP incompatibility. Whether or not it reflects an approach toward a nonregular isometric embedding, when thickness decreases, remains an open question~\cite{marder2007crumpling,sharon2010mechanics}.
The geometric origin of this phenomena was hypothesized to be related to Hilbert’s theorem~\cite{do2016differential}, which rules out any complete surface of constant negative Gaussian curvature in $\mathbb{R}^3$.
It therefore has long been viewed as an exclusive feature of a surface with negative curvature.

Our work overturns this interpretation. By considering growth with \textit{positive} Gaussian curvature we show that horizons may emerge in positively curved geometries. This observation
expands the landscape of geometric frustration to topological frustration in open thin sheets beyond its traditionally understood boundaries.

\begin{figure}[htbp]
\includegraphics{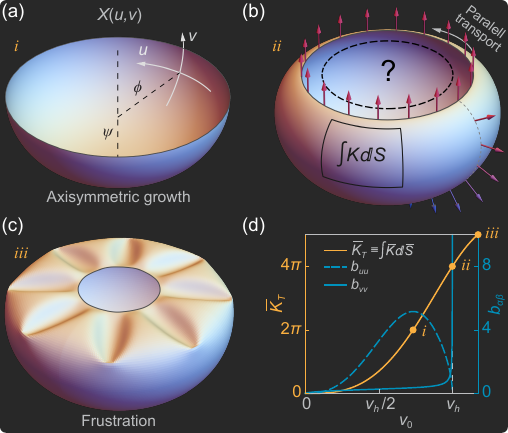}
\caption{\label{fig:Fig1} Emergent frustration in an axisymmetric growing disk with positive Gaussian curvature. (a) Growth of a disk domain endowed with a positive reference Gaussian curvature field $\bar K(v)=K_0 v/v_l$. The total reference curvature $\bar{K}_T = 2\pi$. (b) Onset of a geometric horizon at $\bar K_T=4\pi$, where the boundary normals (arrows) collapse into a common direction, obstructing further smooth isometric extension. (c) For $\bar K_T>4\pi$, symmetry breaking occurs. (d) At the horizon $v_0=v_h$, the principal curvature $b_{vv}$ becomes singular.}
\end{figure}

\textit{First observation: Disk topology—}We start by numerically computing the energy-minimizing shapes of a circular disk endowed with a reference metric of radially increasing positive Gaussian curvature, while growing the system so that the total reference curvature $\bar K_T\equiv\int \bar K\,d\bar S$ crosses $4\pi$ (Fig.~\ref{fig:Fig1}; see SM~\cite{SM2025}\nocite{lee2016fabrication}). For $\bar K_T<4\pi$, the shapes are smooth and axisymmetric. As $\bar K_T\to 4\pi$, the boundary normals collapse toward a common direction and the boundary curvature grows sharply. For $\bar K_T>4\pi$, the minimizing shapes break the axial symmetry, suggesting a geometric horizon at $\bar K_T=4\pi$.
This observation raises a central question: why does an open surface, despite its free boundary, develop a geometric horizon once $\bar K_T$ reaches $4\pi$, and what are the geometric and mechanical consequences of growth beyond this horizon?

\begin{figure}[t]
\includegraphics{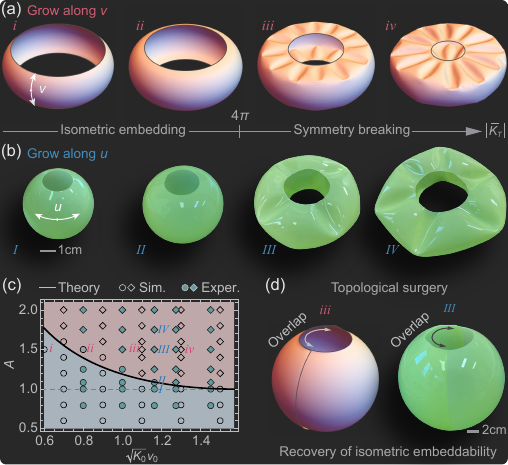}
\caption{\label{fig:Fig2} Pattern formation induced by the isometric frustration, and recovery of isometric embeddability. (a),(b) Symmetry breaking during the
increase of accumulated Gaussian curvature in (a) simulations and (b) experiments. Near $\bar{K}_T = 4\pi$, the edge reaches a geometric horizon; for $\bar{K}_T > 4\pi$, isometric embeddings break and periodic dimples with Pogorelov-like ridge and d-cone emerge. (c) Phase diagram showing the theoretical boundary $v_0=v_h$ (solid curve). Symbols denote simulations (open) and experiments (filled); circles and square indicate isometric and frustrated configurations, respectively. (d) Topological surgery restores smooth isometric embeddability. Cutting $iii$ (a) and \textit{III} (b) along $v$ removes the incompatibility and yields overlapped spherical embeddings. }
\end{figure}

\textit{Growth model and geometry—}To address this question and gain analytical insight, we turn to a tractable model in which a surface grows from a closed ring in an axisymmetric manner while maintaining \emph{constant} Gaussian curvature. The domain, with cylindrical topology, is then defined by the coordinates $(u,v)$ with $u$ being $2\pi$-periodic, and $v$ along the growth direction, thus $\mathcal{M}=[0,2\pi)\times[-v_0,v_0]$ (see Fig.~\ref{fig:Fig2} and End Matter). The symmetry allows us to define $v$ to measure the distance from the equator, thus the target metric induced by this growth protocol is
\begin{equation}
\bar{a}=
\begin{pmatrix}
\Phi^2(v) & 0 \\
0 & 1
\end{pmatrix}\;. \label{eq:refMetric}
\end{equation}
The Gaussian curvature is $\bar{K}=-{\Phi''}/{\Phi}$, and the constant curvature growth protocol with $K_0 = 1/R_0^2$ leads to
$\Phi(v) = \frac{A}{\sqrt{K_0}} \cos(\sqrt{K_0} v)$, with $A = P/(2\pi R_0)$ and $P$ the equator's perimeter. The case $A<1$ resembles a North-American football, while $A=1$ yields a South-American football (see SM~\cite{SM2025}). We are interested in $A>1$, as shown in Fig.~\ref{fig:Fig2}.
The first hint of a horizon appears in the isometric embedding of this sheet as a surface of revolution. Upon searching for a configuration in the form $\mathbf{X}(u,v)=\big(\phi(v)\cos u,\;\phi(v)\sin u,\;\psi(v)\big)$, with $a=\bar{a}$ we find
\begin{eqnarray}
\phi(v)= \frac{A}{\sqrt{K_0}} \cos(\sqrt{K_0} v), \quad 
\psi(v)=\frac{\mathrm{E}(\sqrt{K_0}v\,|\,A^2)}{\sqrt{K_0}}, \label{eq:isometricSol2}
\end{eqnarray}
where $\mathrm{E}(\cdot|\cdot)$ is the elliptic integral of the second kind.
This solution applies equally for positive or negative curvature $K_0$.
An immediate observation is that while for $A\leq 1$ the elliptic integral is well-defined over the whole interval $v \in [-\pi/2,\pi/2]/\sqrt{K_0}$, for $1<A$ it diverges at $v_h = \arcsin(A^{-1})/\sqrt{K_0}$, defined by $\psi'(v_h)=0$. This defines a horizon  along which the unit normal $\hat{\mathbf{n}}$ is constant, and the surface terminates [Fig.~\ref{fig:Fig2}(a), as in Fig.~\ref{fig:Fig1}(b)]. 
On this edge, the principal curvatures exhibit a singular behavior, with $b_{uu}\to 0$ while $b_{vv}\to \infty$ (see End Matter).
We note that this horizon emerges in the configuration, not in the metric, and therefore is not sufficient to exclude the possibility of an asymmetric stretching-free configuration. 

\textit{Isometric incompatibility—}A strong indication for the inability to extend the axisymmetric isometric embedding beyond $v_h$ comes from the notion of rigidifying curves~\cite{audoly1999courbes,audoly2002elastic} and nonlinear isometries developed in~\cite{al2017nonlinear}. In their work it was demonstrated that growth beyond the horizon in the negatively curved pseudosphere is impossible.
Suppose, toward contradiction, that a smooth isometric embedding of the full domain exists even when the growth extends past $v_h$. Then any subdomain of this embedding must also be isometrically embedded. In particular, the restriction to the region $v\in[-v_h,v_h]$ must coincide with the surface-of-revolution isometry obtained above, since the isometry equations admit a unique solution for this metric with the prescribed symmetry. 

Crucially, the uniqueness of the solution reflects the fact that the curve $v=v_h$ is a rigidifying curve. Its normal curvature vanishes, eliminating all regular infinitesimal isometries across it. More severely, along this curve one principal curvature diverges in the $v$ direction for the surface-of-revolution solution. This divergence drives a singularity in the nonlinear isometry equation as well. Thus, similarly to the generic parabolic curves treated in~\cite{al2017nonlinear}, here the blowup of the principal curvature prevents finite-amplitude nonlinear isometries from crossing the curve. 
Consequently, any attempt to continue the isometric embedding from the surface of revolution beyond $v_h$ would require solving the Weingarten equations for a surface~\cite{do2016differential}, whose first and second fundamental forms match those of the surface of revolution on $v\leq v_h$. Yet these equations inherit the same curvature singularity: the coefficient matrix loses ellipticity at $v_h$, and the continuation problem becomes ill-posed. Therefore, the sheet cannot be grown beyond the horizon while preserving the metric. This provides strong evidence for an isometric incompatibility of the growth protocol. It is important to note that this line of reasoning is relevant only for perturbations relative to the surface of revolution. The argument presented here does not exclude the existence of a nonsymmetric isometry and  therefore does not form a complete proof.

While this argument, based on the extrinsic geometry, supports the existence of what we call an isometric incompatibility, it does not expose the intrinsic geometric principle behind its onset. Using the Gauss-Bonnet theorem, we note that regardless of the functional form of $\bar K$, the maximally axisymmetric grown surface reaches a geometric horizon when the accumulated total reference Gaussian curvature satisfies $\bar{K}_T = 4\pi$ (see SM~\cite{SM2025}).

\textit{Beyond the horizon—}To study the mechanical response of the isometrically incompatible spherical-like surfaces, we combine numerical simulations, tabletop experiments and analytical methods. The horizon is located at $v_h = \arcsin(A^{-1})/\sqrt{K_0}$, which suggests that crossing the horizon can be achieved either by fixing the maximal coordinate $v_0$ and increasing $A$, or by fixing $A$ and increasing $v_0$. 

The numerical study is performed within the framework of non-Euclidean elasticity, wherein the elastic energy penalizes for metric and curvature discrepancies ~\cite{efrati2009elastic} 
\begin{equation}
\label{eq:nonEuclideanElasticity}
    \mathcal{E} = \int_{\mathcal{M}} \frac{t}{2} \left\Vert a-\bar{a}\right\Vert^2 + \frac{t^3}{6}  \left\Vert b-\bar{b}\right\Vert^2 \, dS\;.
\end{equation}
Here, $\left\Vert\cdot\right\Vert$ encodes Young's modulus and Poisson ratio (see SM~\cite{SM2025}), and we focus on the thin limit where $a \to \bar{a}$~\footnote{Here, the thin limit is defined by $t$ being much smaller than other characteristic system dimensions, e.g., $t \ll \sqrt{\bar{K}}\, R_0^2$.}. 

In the numerical procedure, we prescribe $\bar{a}$ and $\bar{b}$ and minimize the energy with respect to the surface shape. $\bar{a}$ is taken from Eq.~\eqref{eq:refMetric}, and we run simulations with two types of $\bar{b}$. One that is fully Gauss- and MCP-compatible (i.e., spherical curvature), and one that violates them, $\bar{b}=0$. As shown below and in the SM~\cite{SM2025}, the results are insensitive to this choice.

In Fig.~\ref{fig:Fig2}(a) we show equilibrium configurations as $v_0$ increases. For $v_0 \leq v_h$ solutions are isometric ($a=\bar{a}$, \textit{i}-\textit{ii}), whereas for $v_h<v_0$ symmetry breaks and periodic d-cone patterns emerge~\cite{cerda1998conical,witten2007stress}, indicating the absence of a stretching-free state (\textit{iii}, \textit{iv}).

Interestingly, we find that when a cut is introduced along the $v$ direction, the elastic surface overturns and adopts a multilayered configuration, thereby releasing its residual stresses and restoring the surface-of-revolution solution [Fig.~\ref{fig:Fig2}(d)]. This observation suggests that the geometric effect of the prescribed growth profile is equivalent to inserting an additional azimuthal sector, much like in the classical Volterra construction. In turn, this points to a possible topological character of the resulting frustrated state.

In the experimental study we increase $A$, with a fixed value of $v_0$, via a Volterra-type construction (see SM~\cite{SM2025}). 
We start from casting a spherical shell that corresponds to $A=1$ [Fig.~\ref{fig:Fig2}(b),\textit{I}], then cut it along a meridian and insert an additional wedge with matching geometry to increase $A$. In Fig.~\ref{fig:Fig2}(b),\textit{II} we show a case with $v_0<v_h$, while in \textit{III} and \textit{IV} we observe the frustrated states beyond the horizon with $v_h<v_0$. 

In both simulations and experiments, we analyze the equilibrium configurations preserving or breaking the $u$ symmetry, and identify the transition between isometric embeddings ($a=\bar{a}$) and frustrated ones. This yields the phase diagram in Fig.~\ref{fig:Fig2}(c), parametrized by the dimensionless equator perimeter $A$ and domain size $\sqrt{K_0} v_0$. The transition is cleanly separated by the horizon
$v_h = \arcsin(A^{-1})/\sqrt{K_0}$.

\begin{figure}[t]
\includegraphics{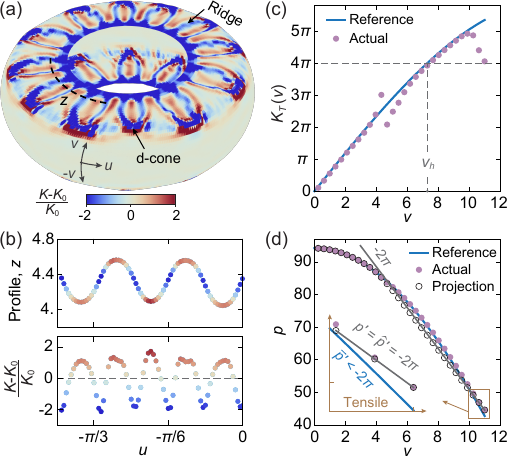}
\caption{\label{fig:Fig3} Posthorizon analysis. (a) Gaussian curvature map showing a nearly flat inner edge preceded by curvature undulations. (b) Cross-section profile and curvature discrepancy, revealing a narrow band of Pogorelov ridges. (c) Accumulated Gaussian curvature along $v$: initially following $\bar{K}_T(v)$ in the interior, but collapses to $4\pi$ at the edge. (d) Perimeter versus $v$: deviations between actual ($p$), reference ($\bar{p}$), and projected ($\hat{p}$) perimeters reflect symmetry breaking, and highlight the tensile nature of the inner edge by $p(v_0) = \hat{p}(v_0)> \bar{p}(v_0)$.}
\end{figure}

\textit{Curvature diagnostics—}Next, we visualize the Gaussian curvature discrepancy $(K-K_0)/K_0$ on the actual configuration [Fig.~\ref{fig:Fig3}(a)]. While regions close to the equator satisfy $K = K_0$, there is an extended frustrated region with spatially oscillating curvature discrepancy. Close to the boundary separating the isometric and frustrated regions, positive and negative peaks localize in a form strongly reminiscent of d-cone structures~\cite{cerda1998conical}. Cross-sectional profiles taken along the azimuthal direction [dashed line in Fig.~\ref{fig:Fig3}(a)] reveal intense  curvature oscillations of the order of $K_0$. This is indicative of Pogorelov ridges that are characterized by localized regions with negative curvature ~\cite{pogorelov1988bendings}.
Both d-cones and Pogorelov ridges are canonical examples of stress-focusing structures in confined thin sheets ~\cite{witten2007stress}. They indicate a qualitative departure from the wrinkle patterns observed along the edges of hyperbolic surfaces [Fig.~\ref{fig:Fig4EM}(a)].  

The integrated Gaussian curvature $K_T(v)$ accumulated along the meridional direction $v$ is plotted in Fig.~\ref{fig:Fig3}(c).  We see that $K_T(v)$ closely follows the reference value $\bar K_T(v)$, until it reaches a critical point where the configuration collapses into a plateau. Interestingly, this critical point precedes the horizon $v_h$ and intrudes back into the isometrically compatible regime, producing a sudden burst in $K_T(v)$. Beyond this point, $K_T(v)$ again tracks $\bar K_T(v)$ and overshoot $4\pi$. However, as the rim is circular and flat $K_T(v)$ relaxes back to the global value of $4\pi$.

A complementary perspective comes from the perimeter evolution along $v$ [Fig.~\ref{fig:Fig3}(d)]. For axisymmetric topologies, critical growth corresponds to a limiting perimeter slope $|p'|_c=2\pi$. When the reference perimeter exceeds this rate ($|\bar p'|>2\pi$), the radius-height slope diverges ($d\phi/d\psi\to\infty$), precluding any smooth axisymmetric embedding and thereby leading to  frustration. In practice, the actual perimeter $p$ closely follows the reference $\bar p$ over most of the domain, showing that even the dimpled regions globally conserve lengths. Locally, however, $p'$ overshoots the $2\pi$ bound, accommodated by symmetry breaking and dimple formation where $p$ deviates from its projection $\hat{p}$. Enlarging the area near the inner plateau reveals $p = \hat{p} >\bar p$ and $p'= \hat{p}'= -2\pi>\bar p'$, showing a flat rim sustained by tensile azimuthal stresses, consistent with a low-order elastic estimate: azimuthal stress switches from compression in the dimples to tension near the rim (see SM~\cite{SM2025}).  

Taken together, these diagnostics establish that elliptic surfaces relieve the isometric incompatibility not by distributed wrinkling as in hyperbolic sheets, but via localized dimples bounded by stress-focusing ridges and d-cones. Despite local excursions and unlike the case of hyperbolic sheets, the total Gaussian curvature $K_T(v_0)$ never exceeds the global bound $4\pi$.

\textit{Discussion—}In summary, we have studied equilibrium configurations of thin circular and annular elastic sheets with positive reference Gaussian curvature. We uncovered a new form of \emph{isometric incompatibility}: a geometric frustration that emerges in open (finite size) sheets even when they are locally compatible. Specifically, we suggest that when the amount of Gaussian curvature, integrated on a disk or annulus equals $4\pi$, the edges of axisymmetric configurations become rigidifying curves. As a result, the smooth axisymmetric embedding cannot be further isometrically extended to contain more than $4\pi$ Gaussian curvature. The mechanical outcome of this geometrical constraint is the development of Pogorelov ridges and d-cones that localize stretching energy. The observed equilibrium configurations are residually stressed even if $\bar{a}$ and  $\bar{b}$ are Gauss- and MCP-compatible. This behavior persists even when the thickness is 3 orders of magnitude smaller than all other length scales in the system. As such, it deviates from the Lewicka-Pakzad scenario, suggesting that there is no accessible $W^{2,2}$ isometric embedding in this regime~\cite{lewicka2011scaling}. 
In light of the work of Ref.~\cite{poznyak1995small}, which considers finite hyperbolic domains and identifies nontrivial isometric embeddings, we emphasize that, since our analysis also concerns finite domains, we cannot rule out the existence of complicated isometric embeddings in the present case as well.
Such configurations are not observed experimentally or in simulations. We note that if exist, they are not accessible perturbatively from the surfaces of revolution, especially in the presence of rigidifying edges.

We have also shown that this frustration has topological characteristics, as it can be removed by inserting meridional or radial cuts [see Fig.~\ref{fig:Fig2}(d)]. Such cuts introduce new, nonrigidifying edges, allowing the system to relax. While our analysis clarifies its mechanical manifestations, the geometric origin of the associated $4\pi$ bound and the onset of rigidifying curves remains to be fully understood.  

The relaxation via cutting resembles mechanisms known in negatively curved surfaces~\cite{poznyak1995small} such as Dini-type surfaces that cannot exist on cylindrical topologies. These observations suggest that, when reconstructing the 3D reference metric in the spirit of~\cite{efrati2009elastic,kupferman2014riemannian}, the corresponding monodromy, which accounts for topological charges in 3D solids, may serve as a topological measure of incompatibility~\cite{kupferman2015metric}. What remains unclear is how the relevant topological charge relates to global properties of $\bar{a}$ and $\bar{b}$, and in particular to the horizon and the maximal integrated curvature beyond which rigidity, and consequently incompatibility, emerge.

Our work also demonstrates the interplay between the intrinsic and extrinsic nature of the isometric incompatibility. We have shown that the horizon not only reflects a limiting total curvature, but also corresponds to a divergence in the perimeter-height relation: as $v\to v_h$, the rate of perimeter growth relative to height, $dp/d\psi$, diverges and the breakdown of axial symmetry is inevitable. The Gauss-Bonnet theorem implies that such divergence should emerge if more than $4\pi$ Gaussian curvature is integrated on an axisymmetric configuration (SM~\cite{SM2025}). It is important to note that $4\pi$ is an upper bound. In annular domains, the divergence of $dp/d\psi$ and consequently, the symmetry breaking, can occur at lower values (SM~\cite{SM2025}). Hyperbolic surfaces can accommodate this divergence through large-amplitude wrinkles and to account for the excess integrated Gaussian curvature, by generating geodesic curvature along the free edge [Fig.~\ref{fig:Fig4EM}]. In contrast, growth in elliptic surfaces form localized dimples indicative of stretching and diverging bending energy density. In addition, the flat circular boundary of such surfaces, for which the integrated geodesic curvature is $-2\pi$, cannot compensate for the excess Gaussian curvature beyond $4\pi$. It therefore shows that the actual metric cannot remain identical to the reference metric while satisfying the Gauss-Bonnet theorem [Fig.~\ref{fig:Fig3}(a)].

The identification of this isometric incompatibility establishes a new organizing principle for the mechanics of thin sheets. It reveals that topological constraints can lead to frustration, even in fully Gauss- and MCP-compatible geometries, thereby forcing shape selection through localized stress-focusing structures rather than distributed wrinkling. This mechanism extends the classical landscape of geometric frustration, and we expect it to guide future theoretical and experimental exploration of topology driven shaping in living tissue and self-morphing systems.

\begin{acknowledgments}
\paragraph*{Acknowledgments—}E.S. acknowledges the support of the USA-Israel Binational Science Foundation, grant no. 2020739 and by the Israel Science Foundation, grant no. 2437/20.
M.M. acknowledges support from the Israel Science Foundation (grant no. 1441/19), and from the Kavli Institute for Theoretical Physics, where part of this work was carried out, under NSF Grant PHY-2309135.
\end{acknowledgments}

\textit{Data availability—}The data that support the findings of this article are not publicly available because the simulation files are too large for hosting in a public repository. The code and data are available from the authors upon reasonable request.

\bibliography{MainText}

\clearpage
\section{End Matter}
\paragraph*{Isometric obstruction beyond $4\pi$—}Let $\mathcal{M}$ be a two-dimensional Riemannian manifold defined on the domain $(u,v) \in [0,2\pi) \times [-v_0, v_0]$, equipped with the metric
\begin{equation}
    ds^2 = \Phi(v)^2 du^2 +dv^2 = a_{\mu\nu} \mathrm{d}x^\mu \mathrm{d}x^\nu
\end{equation}
where $\Phi(v)$ is a smooth, positive function and
\begin{equation}
    a = \begin{pmatrix}
    \Phi(v)^2 & 0 \\
    0 &  1
    \end{pmatrix}\;.
\end{equation}
We consider the following specific case:
\begin{equation}
    \Phi(v) = \frac{A}{\sqrt{K_0}} \cos(\sqrt{K_0} \, v)\;,
\end{equation}
which corresponds to a surface of constant Gaussian curvature $K = K_0$. The total Gaussian curvature associated with this surface is 
\begin{equation}
    K_T  \equiv \int_{\mathcal{M}} K \sqrt{\det{a}} \,\mathrm{d}u\mathrm{d}v = 4\pi A \sin(\sqrt{K_0} \, v_0) \;.
\end{equation}

\begin{figure}[h]
\includegraphics{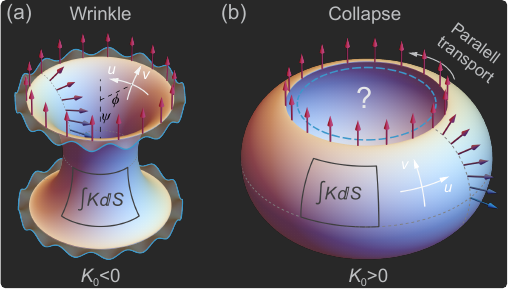}
\caption{\label{fig:Fig4EM} Embedding limitations and singularities in surfaces of constant Gaussian curvature $K_0$. (a) Wrinkling beyond the horizon in a hyperbolic surface ($K_0<0$) ~\cite{sharon2010mechanics,marder2006geometry}. (b) Onset of a horizon in an elliptic surface ($K_0>0$). For both hyperbolic and elliptic surfaces, the horizon forms a rigidifying curve with the normal (indicated by red arrows)  constant on the edge. Note that at the horizon $v_0=v_h$, the total Gaussian curvature reaches $|\bar{K}_T|=4\pi$, while the principal curvatures are singular with $b_{uu}\to 0$ and $b_{vv}\to\infty$.}
\end{figure}

An isometric embedding of this manifold into $\mathbb{R}^3$ can be constructed as a surface of revolution [cf. Fig.~\ref{fig:Fig4EM}(b)],
\begin{equation}
    \mathbf{X}(u,v) = \big( \phi(v)\cos u,\ \phi(v)\sin u,\ \psi(v) \big),
\end{equation}
with
\begin{align}
    \phi(v) &= \frac{A}{\sqrt{K_0}} \cos \left(\sqrt{K_0}\, v\right), \\
    \psi(v) &= \frac{1}{\sqrt{K_0}}\, \mathrm{E}\left(\sqrt{K_0}\, v \,\big|\, A^2\right),
\end{align}
where $\mathrm{E}$ is the incomplete elliptic integral of the second kind.
The corresponding shape operator is
\begin{equation}
    s = a^{-1} b = \begin{pmatrix}
    \kappa_1 & 0 \\
    0 &  \kappa_2
    \end{pmatrix}
    =  
    \sqrt{K_0}\begin{pmatrix}
    f(v)^{-1} & 0 \\
    0 &  f(v)
    \end{pmatrix}\;,
\end{equation}
with
\begin{equation}
    f(v) = \frac{A \cos(\sqrt{K_0} \, v)}{\sqrt{1 - A^2 \sin^2(\sqrt{K_0} \, v)}}\;.
\end{equation}

This form of the second fundamental form reveals two singularities of the configuration. First, regardless of the value of $A$, $f(v) \to 0$ when $v \to  v_s \equiv\pi/(2 \sqrt{K_0})$, with principal curvatures:
\begin{align}
    \kappa_1 &= \sqrt{K_0}\, f(v_0)^{-1}  \to \infty,\\
    \kappa_2 &= \sqrt{K_0}\, f(v_0) \to 0.
\end{align}
However, when $1<A$, which is the case we are interested in, the denominator of $f$ vanishes when $v \to v_h \equiv\sin^{-1}(A^{-1})/\sqrt{K_0}$. In this case 
\begin{align}
     &\kappa_1 = \sqrt{K_0}\, f(v_h)^{-1} \to 0,  \\
     &\kappa_2 = \sqrt{K_0}\, f(v_h)\to \infty,\\
     &b_{uu}\big|_{v_h} \to 0,\quad b_{vv}\big|_{v_h} \to \infty.
\end{align}

We note that $v_s$ forms a singularity not only of the embedding, but also of the metric itself. Contrary to that, $v_{h}$ is a singularity of the embedding, but not of the metric, thus suspected as a horizon (horizons are not true singularity as they can be moved~\cite{poznyak1995small,sharon2010mechanics}). 
To demonstrate that no smooth isometric embedding can be generated by extending the surface to $v_h < v_0$, we proceed by contradiction. Assume such an embedding exists. Then its restriction to the subdomain $v \in [-v_h, v_h]$ must coincide (up to rigid motion) with the known embedding described above. However, according to Ref.~\cite{al2017nonlinear}:

\begin{itemize}
    \item The boundaries at $v = \pm v_h$ form rigidifying curves since the normal curvature $\kappa_N = \kappa_1 = 0$, so linear isometries do not exist.
    \item The second fundamental form diverges: $\kappa_2 \to \infty$, thus excluding nonlinear isometries as well.
    \item The Weingarten equations become singular and cannot be integrated across $v = v_h$, thus excluding the possibility of an isometry beyond the horizon.
\end{itemize}

It is important to note that this argument lacks a rigorous proof for one step: based on the absence of nonlinear isometries we hypothesize that higher order nonlinear isometries do not exist, and therefore the surface is rigid. In that case, from the singularity of Weingarten equations the normal field cannot be extended and the surface cannot be continued beyond $v = v_h$.

\end{document}


\title{Supplemental Material for "Isometric Incompatibility in Growing  Elastic Sheets"}

\author{Yafei Zhang}
\author{Michael Moshe}%
\email{michael.moshe@mail.huji.ac.il}
\author{Eran Sharon}
\email{erans@mail.huji.ac.il}

\affiliation{Racah Institute of Physics, The Hebrew University of Jerusalem, Jerusalem, 9190401, Israel.}%

\maketitle

\tableofcontents
\vspace{1em}

\newpage

\section{Isometric Embeddings with Constant Gaussian Curvature}

In this section, we revisit the classical isometric embeddings of surfaces with constant Gaussian curvature~\cite{do2016differential}, which serve as the geometric foundations for the analysis in the main text. We derive explicit parametrizations for both negatively and positively curved surfaces of revolution, and classify the corresponding families of embeddings according to their boundary conditions and singularity structures. These canonical examples illustrate how curvature sign and topology dictate the possible terminations of smooth embeddings in $\mathbb{R}^3$.

We consider a reference domain with cylindrical topology endowed with an axisymmetric reference metric
\begin{equation}
\label{eq:refMetric}
\bar{a}=\begin{pmatrix}
        \Phi^2(v) & 0 \\
        0 & 1 
        \end{pmatrix},
\end{equation}
defined on the parameter domain $\mathcal{M}= [0,2\pi)\times [-v_0, v_0]$.  
The corresponding Gaussian curvature is
\begin{equation}
\label{eq:refK}
    \bar{K}=-\frac{\Phi''}{\Phi}.
\end{equation}

The surface of revolution in $\mathbb{R}^3$ is parameterized as
\begin{equation}
\label{eq:X}
    \mathbf{X}(u,v)=(\phi(v)\cos u,\;\phi(v)\sin u,\;\psi(v)),
\end{equation}

with actual metric
\begin{equation}
    a=\begin{pmatrix}
      \phi^{2}(v)  & 0 \\
       0 & \phi'^2(v)+\psi'^2(v)
       \end{pmatrix}.
\end{equation}

Requiring an isometric embedding with constant Gaussian curvature $K_0$ (i.e.\ $\bar K = K_0$ and $\bar a=a$) leads to
\begin{align}
&\phi'' + K_0 \phi =0, \label{eq:generalEQ1}\\
&\psi(v) =\int_0^{v}\sqrt{\,1-\phi'^2(v)}\,dv. \label{eq:generalEQ2}
\end{align}
The resulting surfaces depend both on the sign of $K_0$ and on the imposed boundary conditions.

It is straightforward to show that
\begin{equation}
\label{eq:p'/psi'}
    \frac{d p}{d \psi} = 2\pi \frac{d \phi}{d \psi} 
    = 2\pi \frac{\phi'(v)}{\sqrt{1-\phi'^2(v)}},
\end{equation}
where $p(v)$ denotes the perimeter of the axisymmetric configuration at the meridional coordinate $v$.

The total reference Gaussian curvature can then be written as
\begin{equation}
\label{eq:RefKT}
    \bar{K}_T
    = \int_{\mathcal{M}} \bar{K}(v)\sqrt{\det\bar{a}}\;dudv
    = 2\pi\big(\phi'(-v_0)-\phi'(v_0)\big).
\end{equation}

\subsection{$K_0<0$: Surfaces of revolution with constant negative Gaussian curvature}

In this case the general solution of Eq.~(\ref{eq:generalEQ1}) is
\begin{equation}
    \phi(v) = c_1 \cosh(\sqrt{-K_0}\,v) + c_2 \sinh (\sqrt{-K_0}\,v),
    \label{eq:generalSolN}
\end{equation}
with constants $c_1,c_2$ determined by boundary conditions.  
Different choices of $(\phi(0),\phi'(0))$ yield three canonical families of embeddings, each with distinct singularity structure.

\subsubsection{Pseudosphere}

Imposing equatorial data
\[
\phi(0)=\tfrac{A}{\sqrt{-K_0}},\qquad \phi'(0)=A,
\]
one selects the decaying branch $\phi(v)=(A/\sqrt{-K_0}) e^{-\sqrt{-K_0} v}$.  
With the substitution $e^{\sqrt{-K_0}v}=\cosh v_1$, Eqs.~\eqref{eq:generalEQ1}--\eqref{eq:generalEQ2} yield
\begin{align}
   \phi(v_1) &= \tfrac{A}{\sqrt{-K_0}}\mathrm{sech} v_1,\\
   \psi(v_1) &= \tfrac{A}{\sqrt{-K_0}}(v_1-\tanh v_1).
\end{align}
This is the classical \emph{pseudosphere} (Beltrami surface), terminating in two point-like singularities at the vertical axis as $v_1\to\pm\infty$ (Fig.~\ref{figS:FigS1}a).

\subsubsection{Spindle-like pseudospherical surface}

For axis data
\[
\phi(0)=0,\qquad \phi'(0)=A,\quad 0<A<1,
\]
Eqs.~\eqref{eq:generalEQ1}--\eqref{eq:generalEQ2} give
\begin{align}
   \phi(v) &= \tfrac{A}{\sqrt{-K_0}}\sinh(\sqrt{-K_0}\,v),\\
   \psi(v) &= \frac{i\sqrt{(A^2-1)K_0}\,\mathrm{E}\!\left(i \sqrt{-K_0}v\;\bigg|\;\tfrac{A^2}{A^2-1}\right)}{K_0}.
\end{align}
This configuration exhibits a \emph{point-like conical singularity} at the axis ($v=0$, deficit angle $\delta=2\pi(1-A)$). In addition, at finite $v=v_s$ defined by $\cosh(\sqrt{-K_0} v_s)=1/A$, the surface terminates at a \emph{circular singularity} where $\psi'(v_s)=0$. Thus the spindle-like pseudospherical surface interpolates between an axial point singularity and a rim singularity (Fig.~\ref{figS:FigS1}b).

\subsubsection{Bulge-like pseudospherical surface}

For symmetric equatorial conditions
\[
\phi(0)=\tfrac{A}{\sqrt{-K_0}},\qquad \phi'(0)=0,
\]
Eqs.~\eqref{eq:generalEQ1}--\eqref{eq:generalEQ2} yield
\begin{align}
   \phi(v) &= \tfrac{A}{\sqrt{-K_0}}\cosh(\sqrt{-K_0}\,v),\\
   \psi(v) &= -\tfrac{i}{\sqrt{-K_0}}\,\mathrm{E}\!\left(i \sqrt{-K_0}v\;\big|\;-A^2\right).
\end{align}
This \emph{bulge-like} pseudospherical surface is symmetric about the equator and ends at two circular singularities located at finite meridional distance from the axis (Fig.~\ref{figS:FigS1}c).

\subsection{$K_0>0$: Surfaces of revolution with constant positive Gaussian curvature}

\begin{figure*}[t]
\centering
\includegraphics[width=1.0\linewidth]{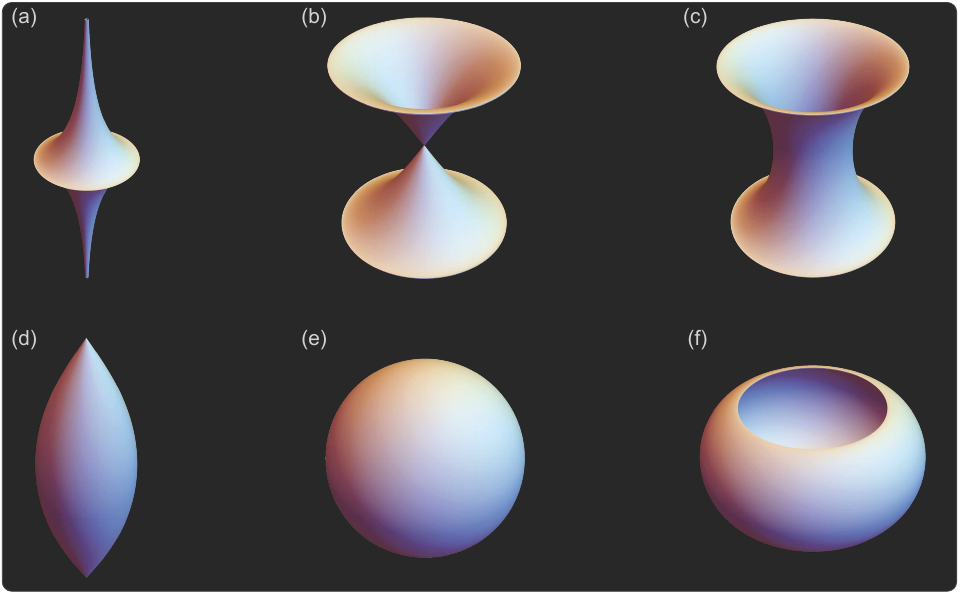}
\caption{\label{figS:FigS1} Isometric embeddings of axisymmetric surfaces with constant Gaussian curvature. (a–c) Constant negative curvature: (a) pseudosphere with two point singularities on the axis; (b) spindle-like pseudospherical surface with a conical singularity at the axis and two circular singularities at finite radius; (c) bulge-like pseudospherical surface with two circular singularities at the edges. (d–f) Constant positive curvature: (d) spindle surface with two conical singularities at the poles; (e) classical sphere, a complete smooth embedding; (f) bulge surface with two circular singularities at the edges.}
\end{figure*}

Here the general solution of Eq.~\eqref{eq:generalEQ1} reads
\begin{equation}
    \phi(v) = c_1 \cos(\sqrt{K_0}\,v) + c_2 \sin(\sqrt{K_0}\,v),
    \label{eq:generalSolP}
\end{equation}
with $c_1,c_2$ fixed by boundary data.  
Under the natural symmetric condition
\[
\phi(0)=\tfrac{A}{\sqrt{K_0}},\qquad \phi'(0)=0,
\]
one obtains
\begin{align}
\phi(v) &= \tfrac{A}{\sqrt{K_0}}\cos(\sqrt{K_0}\,v), \label{eq:isometricSol1}\\
\psi(v) &= \tfrac{1}{\sqrt{K_0}}\,E\!\left(\sqrt{K_0}v\,\big|\,A^2\right).\label{eq:isometricSol2}
\end{align}
Depending on $A$, this encompasses three canonical cases:

\paragraph*{Spindle surface}(North-American football). For $A<1$, Eqs.~\eqref{eq:isometricSol1}--\eqref{eq:isometricSol2} describe a spindle with two cuspidal point singularities along the axis (Fig.~\ref{figS:FigS1}d).  

\paragraph*{Classical sphere}(South-American football). For $A=1$, the embedding reduces to the classical sphere of radius $1/\sqrt{K_0}$ (Fig.~\ref{figS:FigS1}e).  

\paragraph*{Bulge surface.} For $A>1$, one obtains a bulge-type surface that terminates at two circular singularities at finite meridional distance (Fig.~\ref{figS:FigS1}f).  

\section{Material and Fabrication}

“Growing” a surface in the experiment according to the reference metric~\eqref{eq:refMetric}, i.e., increasing $v_0$ or $A$, is much more challenging than in the simulation. Note that rescaling $u \equiv \theta /A $ maps the reference metric~\eqref{eq:refMetric} to that of a sphere with $A=1$ and $K=K_0$, so cases $A\neq 1$ in~\eqref{eq:refMetric} can be achieved via a Volterra-type construction that inserts (for $1<A$) or removes (for $A<1$) an angular wedge. Accordingly, we experimentally "grow" a surface in $u$-direction, starting from casting a spherical shell which corresponds to $A=1$. 

\begin{figure*}[t]
\centering
\includegraphics[width=1.0\linewidth]{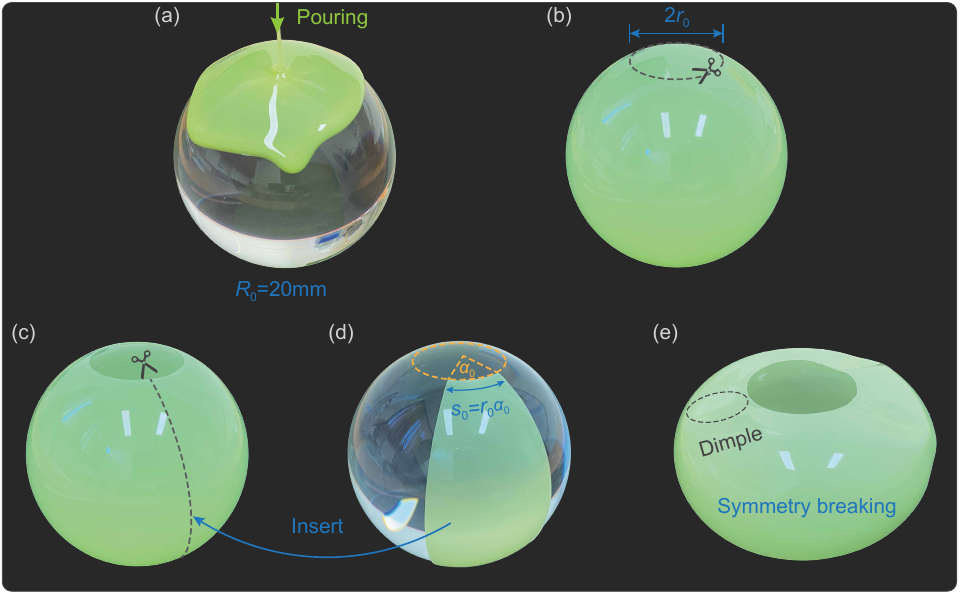}
\caption{\label{figS:FigS2} 
Experimental protocol for “growing’’ a spherical surface. 
(a) VPS solution is poured onto a glass ball of radius $R_0=20$\,mm and cured. 
(b) Removing two spherical caps of radius $r_0$ produces a spherical zone. 
(c) The zone is cut along a meridian. 
(d) A spherical sector of swivel angle $\alpha_0$ and arc length $s_0=r_0\alpha_0$ is inserted. 
(e) Gluing the zone and sector together increases the effective perimeter; when $\bar{K}_T>4\pi$, rotational symmetry breaks and dimples form.}
\end{figure*}

We fabricate thin elastic spherical shells using a coating–drainage–curing method introduced by Lee \textit{et al.}~\cite{lee2016fabrication}. A vinylpolysiloxane (VPS) solution (Elite Double 32, Zhermack) is prepared by mixing base and curing agent at a 1:1 weight ratio for 15\,s at room temperature (20\,$^\circ$C). The mixture is rapidly poured onto a glass ball of radius $R_0=20$\,mm (Fig.~\ref{figS:FigS2}a), whose surface is pre-treated with a release agent (Ease Release 200, Mann) to facilitate demoulding. During drainage, a soft plastic brush is used to distribute the liquid uniformly across the sphere. After curing for 40\,min, the VPS coating is detached, yielding a thin spherical shell conforming to the glass ball (Fig.~\ref{figS:FigS2}b).  

A spherical zone is obtained by removing two concentric spherical caps of radius $r_0$ from the poles (Fig.~\ref{figS:FigS2}b). Its total Gaussian curvature is  
\begin{equation}
\label{eq:Kspherezone}
    K_{sz} = \int_{sz} K_0\,dS = 4\pi \sin\!\left(\tfrac{v_0}{R_0}\right) 
    = 4\pi \sin(\sqrt{K_0}v_0) 
    = 4\pi \sqrt{1-\tfrac{r_0^2}{R_0^2}}, 
\end{equation}
where $v_0 = R_0 \arccos(r_0/R_0)$ is the effective meridional extent, and $K_0=R_0^{-2}$ is the constant Gaussian curvature.

To implement growth in the azimuthal ($u$) direction, the spherical zone is cut along a meridian (Fig.~\ref{figS:FigS2}c), and a spherical sector with swivel angle $\alpha_0$ and arc length $s_0=r_0\alpha_0$ is inserted (Fig.~\ref{figS:FigS2}d). Using the glass sphere as a support, the zone and the sector are bonded together with fresh VPS solution. Removing the mold yields a grown surface whose reference total curvature is  
\begin{equation}
\label{eq:KTspherezone}
    \bar{K}_T = 4\pi \sin(\sqrt{K_0}v_0)\left(1+\tfrac{\alpha_0}{2\pi}\right) 
    = 4\pi A \sin(\sqrt{K_0}v_0), 
\end{equation}
where $A \equiv 1+\alpha_0/2\pi>1$ is the perimeter factor.  On the contrary, removing a sector (corresponding to a negative swivel angle $\alpha_0$) from the spherical zone, and then directly gluing the two meridian boundaries together, will result in a surface with $A=1+\alpha_0/2\pi <1$.

This procedure realizes controlled "growth" of a spherical zone with constant curvature $K_0$. When $A > \sin^{-1}(\sqrt{K_0}v_0)$ (equivalently $\alpha_0>2\pi(\sin^{-1}(\sqrt{K_0}v_0)-1)$), the total reference curvature exceeds the threshold $\bar{K}_T=4\pi$, symmetry is broken, and dimples emerge (Fig.~\ref{figS:FigS2}e).  

\begin{figure*}[t]
\centering
\includegraphics[width=1.0\linewidth]{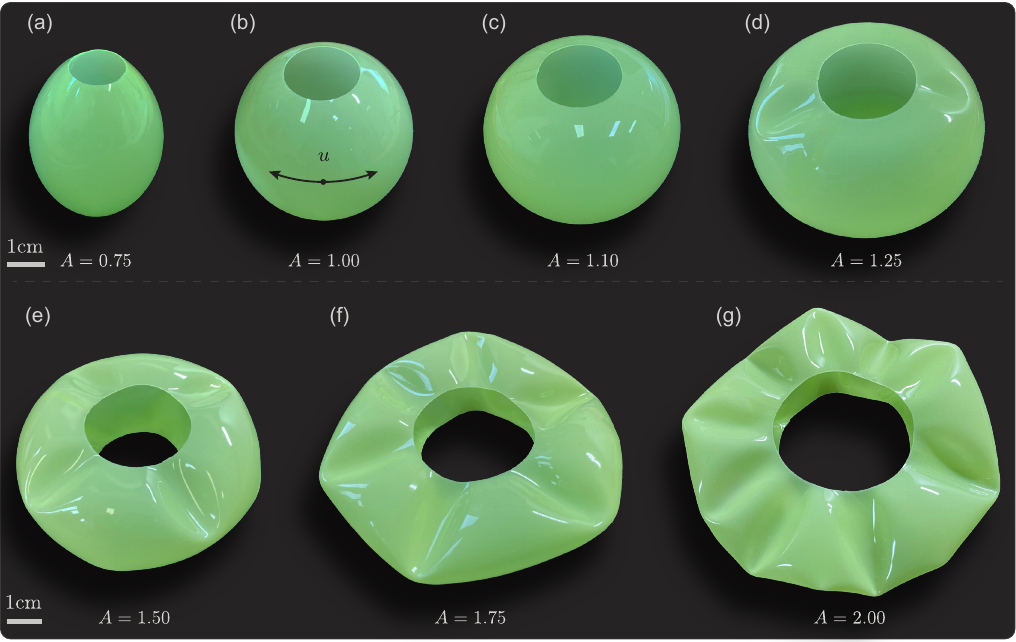}
\caption{\label{figS:FigS3} 
Experimental realization of the growing spherical surfaces. (a–d) Sequential growth of a spherical surface along the $u$-direction within the isometric regime ($\bar K_T \le 4\pi$, $A \le A_h$), where smooth spherical configurations are obtained.  Configuration (a) corresponds to a deficit--growth case ($A < 1$) relative to the initial spherical zone (b) with $A = 1$; all other configurations correspond to wedging growth ($A > 1$). Specifically, configuration (c) represents the near--critical state in which the edge approaches the geometric horizon.
(e--g) Beyond the isometric limit ($\bar{K}_T > 4\pi,\; A > A_h$), symmetry breaking occurs and the surfaces develop frustrated configurations with dimples. The number of dimples increases with $A$, reflecting the progressive build--up and release of geometric frustration during growth.
The initial geometric parameters of the spherical zone (b) are $R_0 = 25\,\mathrm{mm}$, $r_0 = 10\,\mathrm{mm}$, and $t=0.22~\mathrm{mm}$ (upper hemisphere thickness).}
\end{figure*}

\section{The Geometry of Isometric Embedding}

For a surface of revolution $\mathbf{X}\left(u,v\right)=\left(\phi(v)\cos{u}, \phi(v)\sin{u}, \psi(v)\right)$, the first and second fundamental forms read,

\begin{subequations}\label{eq:SoRa}
\begin{align}
a_{\alpha\beta}=\partial _{\alpha }\mathbf{X} \cdot \partial _{\beta }\mathbf{X} &=
\begin{pmatrix}
\phi^{2} & 0\\
0 & \phi'^2+\psi'^2
\end{pmatrix}\label{eq:SoRa:a}\\
&=
\begin{pmatrix}
\frac{A^2}{K_0}\cos^2(\sqrt{K_0}v) & 0\\
0 & 1
\end{pmatrix}\label{eq:SoRa:b}
\end{align}
\end{subequations}

\begin{subequations}\label{eq:SoRb}
\begin{align}
b_{\alpha \beta }&=\hat{\mathbf{N}} \cdot \partial _{\alpha } \partial_{\beta }\mathbf{X}=\begin{pmatrix}
      -\frac{\phi\psi'}{\sqrt{\phi'^2+\psi'^2}}  & 0 \\
       0 & \frac{\phi''\psi'-\phi'\psi''}{\sqrt{\phi'^2+\psi'^2}} 
       \end{pmatrix}\label{eq:SoRb:a}\\
&=- \cos{(\sqrt{K_0}v)} \begin{pmatrix}
     \frac{A \sqrt{1-A^2\sin^2{(\sqrt{K_0}v)}}}{\sqrt{K_0}}  & 0 \\
       0 & \frac{A \sqrt{K_0}}{\sqrt{1-A^2\sin^2{(\sqrt{K_0}v)}}}
       \end{pmatrix}\label{eq:SoRb:b}
\end{align}
\end{subequations}

with the unit normal vector
\begin{subequations}\label{eq:SoRn}
\begin{align}
\hat{\mathbf{n}} &= \frac{\partial_{u}\mathbf{X} \times \partial_{v}\mathbf{X}}{|\partial_{u}\mathbf{X} \times \partial_{v}\mathbf{X}|} =\frac{1}{\sqrt{\phi'^2+\psi'^2}}(\cos(u)\psi',\sin(u)\psi',-\phi')\label{eq:SoRn:a}\\
&={\sqrt{1-A^2\sin^2{(\sqrt{K_0}v)}}}\left(\cos(u),\sin(u),\frac{A\sin(\sqrt{K_0}v)}{{\sqrt{1-A^2\sin^2{(\sqrt{K_0}v)}}}}\right)\label{eq:SoRn:b}
\end{align}
\end{subequations}

The Weingarten equations can be written as,
\begin{equation}
    \partial_{\alpha}\hat{\mathbf{n}} = b_{\alpha}^{\cdot\beta} \partial_{\beta}\mathbf{X}.
\end{equation}

\begin{figure*}[t]
\centering
\includegraphics[width=1.0\linewidth]{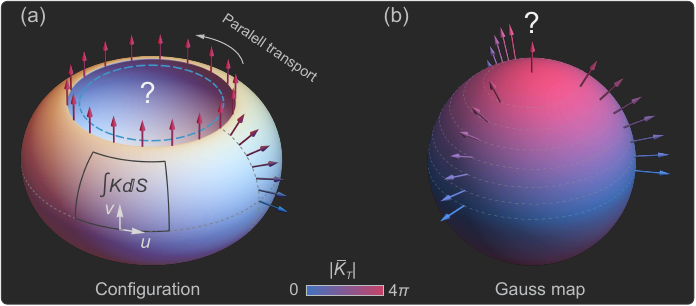}
\caption{\label{figS:FigS4} {The emergence of a geometric horizon and its Gauss map.} 
(a) Schematic illustration of the configuration near the embedding limit, showing the distribution of surface normals (arrows) and the cumulative Gaussian curvature $\int K\,dS$ over the surface. As the total curvature approaches the limit $4\pi$, the normals along the rim become parallel and collapse into a single direction, forming a geometric horizon where the Gauss map degenerates. (b) Corresponding Gauss map on the unit sphere. As curvature accumulates, the image of the surface under the Gauss map progressively expands over $S^2$, eventually filling the entire unit sphere. At the same time, the image of the rim converges to a single point, reflecting the collapse of all boundary normals into one direction. This degeneration of the Gauss map marks the termination of smooth isometric embeddings in $\mathbb{R}^3$.}
\end{figure*}

\section{Non-Euclidean Elasticity}

To study the isometric embeddings and the pattern formation after the onset of symmetry breaking (isometric embedding absence), we employ the theoretical framework of \textit{non-Euclidean elasticity}~\cite{efrati2009elastic}. 
This framework generalizes classical thin-shell elasticity to materials whose natural, stress-free configuration cannot be embedded in three-dimensional Euclidean space $\mathbb{R}^3$.

\paragraph*{\textbf{Reference geometry and actual configuration}.}
In this framework, a thin sheet is described by two pairs of geometric fields: the \emph{reference metric} $\bar{a}_{\alpha\beta}$ and \emph{reference curvature} $\bar{b}_{\alpha\beta}$,
which encode the intrinsic distances and natural bending of the material in the absence of external constraints,
and the corresponding \emph{actual metric} $a_{\alpha\beta}$ and \emph{actual curvature} $b_{\alpha\beta}$,
which describe the realized configuration in $\mathbb{R}^3$.
When $(\bar{a},\bar{b})$ satisfy the Gauss–Mainardi–Codazzi–Peterson (GMCP) equations,
they correspond to a surface that can be smoothly embedded in $\mathbb{R}^3$~\cite{do2016differential}:
\begin{equation}
\label{eq:GMCP}
K({\bar{a}}) = \det(\bar{a}^{-1}\bar{b}), \quad
    \bar{\nabla}_{\alpha}\bar{b}_{\beta\gamma}-\bar{\nabla}_{\beta}\bar{b}_{\alpha\gamma} =0.
\end{equation}
where 
\begin{equation}
    K(\bar{a}) = \frac{1}{2}\bar{a}^{\alpha\beta}(\partial_{\gamma}\Gamma^{\gamma}_{\alpha\beta} -\partial_{\beta}\Gamma^{\gamma}_{\gamma\alpha}+\Gamma^{\gamma}_{\gamma\delta}\Gamma^{\delta}_{\alpha\beta}-\Gamma^{\gamma}_{\alpha\delta}\Gamma^{\delta}_{\gamma\beta}),
\end{equation}
the covariant derivative of a second-order tensor $T$,
\begin{equation}
    \bar{\nabla}_\alpha T_{\beta\gamma} =\partial_{\alpha} T_{\beta\gamma}-\Gamma^{\delta}_{\alpha\beta}T_{\delta\gamma}-\Gamma^{\delta}_{\alpha\gamma}T_{\beta\delta},
\end{equation}
and the Levi-Civita connection,
\begin{equation}
    \Gamma_{\alpha\beta}^{\gamma}=\frac{1}{2}\bar{a}^{\gamma\delta}(\partial_{\alpha}\bar{a}_{\beta\delta}+\partial_{\beta}\bar{a}_{\alpha\delta}-\partial_{\delta}\bar{a}_{\alpha\beta}).
\end{equation}

However, in general, inhomogeneous growth, swelling, or plastic deformation
may prescribe incompatible $(\bar{a},\bar{b})$ that violate GMCP compatibility,
so that no configuration can simultaneously satisfy $a=\bar{a}$ and $b=\bar{b}$.
This geometric incompatibility gives rise to residual stresses and spontaneous shape formation.

\paragraph*{\textbf{Elastic energy}.}
Following the Kirchhoff–Love hypothesis, 
the total elastic energy of a thin sheet of thickness $t$ can be expressed as the sum of
stretching and bending contributions that penalize deviations from the reference geometry~\cite{efrati2009elastic}:
\begin{equation}
\mathcal{E} = \int_{\mathcal{M}}
\Big[
t\,\|a-\bar{a}\|^2 
+ \frac{t^3}{3}\,\|b-\bar{b}\|^2
\Big]
\,dS,
\label{eq:NonEuElasticityEnergy}
\end{equation}
and $\|\cdot\|^2 = \frac{Y\nu}{8(1-\nu^2)} \mathrm{Tr}^2[\bar{a}^{-1}(\cdot)]+\frac{Y}{8(1+\nu)}\mathrm{Tr}[\bar{a}^{-1}(\cdot)]^2$ defines the elastic norm in the tangent space~\cite{efrati2009elastic}, where $Y$ and $\nu$ are the Young’s modulus and Poisson’s ratio, respectively.
The first term quantifies in-plane stretching (metric deviation), while the second term accounts for out-of-plane bending (curvature deviation).

\paragraph*{\textbf{Incompatibility and minimization principle}.}
For a geometrically compatible reference geometry, the sheet can achieve a zero-energy isometric embedding ($a=\bar{a}$, $b=\bar{b}$).
When $(\bar{a},\bar{b})$ are incompatible, no such configuration exists, 
and the actual shape $(a,b)$ minimizes the energy~\eqref{eq:NonEuElasticityEnergy}
subject to the GMCP constraints~\eqref{eq:GMCP}, i.e., $K({a}) = \det(a^{-1}b)$ and $\nabla_{\alpha}b_{\beta\gamma}-\nabla_{\beta}b_{\alpha\gamma} =0$, which guarantee that the actual geometry $a,b$ correspond to a valid immersion in $\mathbb{R}^3$.
This constrained minimization captures the competition between stretching and bending energies
that governs morphogenesis, buckling, and pattern formation in thin elastic media.

\paragraph*{\textbf{Thin-limit approximation}.}
The sheet thickness $h$ plays a crucial role in this competition:
for extremely thin sheets ($t\sqrt{K_0}\!\ll\!1$), 
the stretching term dominates, and the system minimizes its energy by enforcing $a=\bar{a}$ if possible,
i.e., by approaching an isometric embedding of the reference metric.
At finite $h$, however, bending and stretching become comparable, 
and the system may develop localized structures such as wrinkles, dimples, or ridges to partially relax the incompatibility.
Beyond the isometric embeddings, these equilibria can be numerically obtained by minimizing Eq.~\eqref{eq:NonEuElasticityEnergy}
using finite element methods under the GMCP compatibility constraints.

\section{Numerical Computations}
\subsection{Finite Element Simulations}

\begin{figure*}[t]
\includegraphics{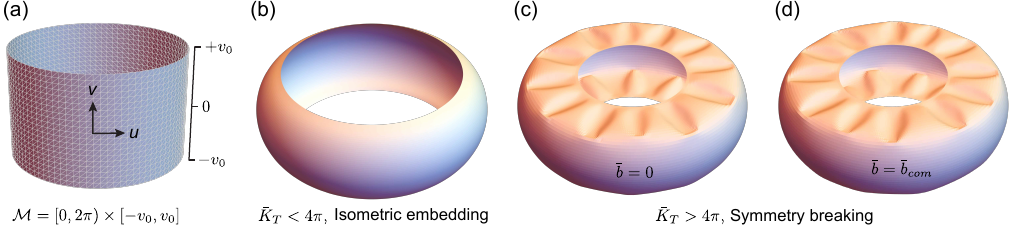}
\caption{\label{figS:FigS5}
Reference domain, isometric embedding, and symmetry breaking. 
(a) A meshed cylindrical parameter domain $\mathcal{M} = [0,2\pi)\times[-v_0,v_0]$ used to assign the reference metric $\bar{a}$ and reference curvature $\bar{b}$. (b) Isometric embedding of the reference metric when the total reference Gaussian curvature satisfies $\bar K_T = 3.86\pi < 4\pi$. 
The surface is smoothly embedded in $\mathbb{R}^3$ and retains full rotational symmetry. (c,d) When $\bar K_T = 5.78\pi > 4\pi$, the isometric embedding constraint can no longer be globally satisfied, leading to a symmetry-breaking instability that generates a dimpled configuration relieving the geometric frustration. 
The parameters in (b,c) are $K_0 = 1/100$, $A = 1.5$, $t = 0.02$, $\bar b = 0$, and $v_0 = 7$ and 11, respectively. 
To verify the robustness of this result, we further simulated the same system but with a compatible reference curvature tensor $\bar{b} = \bar{b}_{com}$ in Eq.~\eqref{eq:comGMCP}. Combined with $\bar a$ in Eq.~\eqref{eq:abar-bbar}, this choice strictly satisfies the Gauss--Mainardi--Codazzi--Peterson (GMCP) equations, yet produces an identical frustrated configuration (d). This confirms that the observed frustration is not due to the slight violation of the Gauss equation and justifies the simplification $\bar b = 0$ in the thin-sheet limit.
}
\end{figure*}

We numerically study the shape evolution of axisymmetric surfaces using an in-house finite element code. Our reference domain is a two-dimensional manifold with cylindrical topology, $\mathcal{M} = [0, 2\pi) \times [-v_0, v_0]$.  
Given the height $2v_0$ and radius $R_0 = 1/\sqrt{K_0}$, we discretize the cylindrical domain into triangular elements (Fig.~\ref{figS:FigS5}a). In the simulation, the reference geometry is set as,
\begin{equation}
\label{eq:abar-bbar}
\bar{a} =\begin{pmatrix}
      \frac{A^2}{K_0}\cos^2{(\sqrt{K_0}v)}  & 0 \\
       0 & 1
       \end{pmatrix},\quad \bar{b} = 0.
\end{equation}
The initial configuration is chosen as a slightly curved cylindrical surface,  
\begin{equation}
\label{eq:initialconf}
X_0(u,v) = 
\big(\alpha A R_0 \cos\tfrac{\beta v_0}{\alpha A R_0} \cos u,\;
      \alpha A R_0 \cos\tfrac{\beta v_0}{\alpha A R_0} \sin u,\;
    \alpha A R_0 \sin\tfrac{\beta v_0}{\alpha A R_0}\big),
\end{equation}
where $\alpha, \beta$ are control parameters that define the initial curvature ($\alpha = 1.05$, $\beta = 3.0$ in all simulations). The normalized thickness is set as a small value i.e., $t\sqrt{K_0} \leq 0.002$. The material properties are set as Young’s modulus $Y = 1000$ and Poisson’s ratio $\nu = 0.3$.  
The convergence of the simulations is ensured by employing more than $5\times10^4$ triangular elements. This number is determined through systematic convergence tests, in which mesh densities ranging from $2\times10^4$ to $1.3\times10^5$ elements were examined. All simulations are performed under quasi-static conditions to minimize numerical artifacts, and the final equilibrium configurations are obtained by minimizing the total elastic energy. Inputting Eqs.~\eqref{eq:abar-bbar} and~\eqref{eq:initialconf} into the code and minimizing the total energy Eq.~\eqref{eq:NonEuElasticityEnergy} subject to the GMCP constraint yield the equilibrium configurations (Fig.~\ref{figS:FigS5}b-c). Note that all finite element simulations are conducted under free boundary conditions, without any external constraints.

Since we mainly focus on the small-thickness regime ($t\sqrt{K_0}\ll1$), we assume $\bar b = 0$ in most simulations for simplicity, without loss of generality. To verify that this approximation does not affect the results, in Fig.~\ref{figS:FigS5}d,  we additionally perform simulations with a compatible reference second fundamental form
\begin{equation}
\label{eq:comGMCP}
    \bar{b}_{com} = \begin{pmatrix}
      \frac{A^2}{\sqrt{K_0}}\cos^2{(\sqrt{K_0}v)}  & 0 \\
       0 & \sqrt{K_0}
       \end{pmatrix}.
\end{equation}
Together with the reference metric $\bar a$ in Eq.~\eqref{eq:abar-bbar}, this choice ensures that the reference geometry strictly satisfies the Gauss--Mainardi--Codazzi--Peterson (GMCP) compatibility conditions in Eq.~\eqref{eq:GMCP}, during the whole growth. The resulting configuration in Fig.~\ref{figS:FigS5}d remains identical to the frustrated state shown in Fig.~\ref{figS:FigS5}c. This confirms the core argument of the paper---that frustration persists even when the reference geometry is fully GMCP-compatible---and validates the simplification $\bar b = 0$ in the thin-sheet limit.

\subsection{Numerical Solutions}

We assume that the resulting configuration takes the form of a surface of revolution for any domain size: $\mathbf{X}(u,v)=\big(\phi(v)\cos{u},\, \phi(v)\sin{u},\, \psi(v)\big)$. For $v_0>v_h$, these axisymmetric numerical solutions represent metastable states rather than the minimal-energy configurations. Nevertheless, their stress distributions provide useful insight into the formation and mechanical features of the rigidifying curve.
The corresponding first and second fundamental forms are obtained from Eqs.~\eqref{eq:SoRa:a} and~\eqref{eq:SoRb:a}, respectively.  
The reference forms are prescribed by Eqs.~\eqref{eq:abar-bbar}.  
The in-plane strain and stress tensors are then expressed as
\begin{equation}
\label{eq:strain-stress}
    u_{\alpha\beta}=\tfrac{1}{2}(a-\bar{a})_{\alpha\beta}, 
    \qquad 
    \sigma^{\alpha\beta} = \mathcal{A}^{\alpha\beta\gamma\delta}u_{\gamma\delta},
\end{equation}
where $\mathcal{A}^{\alpha\beta\gamma\delta}=\frac{\nu Y}{1-\nu^2}
\big(\bar{a}^{\alpha\beta}\bar{a}^{\gamma\delta}
+\tfrac{1-\nu}{2\nu}(\bar{a}^{\alpha\gamma}\bar{a}^{\beta\delta}
+\bar{a}^{\alpha\delta}\bar{a}^{\beta\gamma})\big)$ 
is the elastic tensor that encodes the material’s isotropic constitutive law.

Substituting Eqs.~\eqref{eq:SoRa:a}, \eqref{eq:SoRb:a}, \eqref{eq:abar-bbar}, and~\eqref{eq:strain-stress} into the total elastic energy functional~\eqref{eq:NonEuElasticityEnergy}, we can express the strain energy and the stress function as
\begin{align}
    \mathcal{E} &= \mathcal{E}[\phi(v),\phi'(v),\phi''(v),\, \psi(v),\psi'(v),\psi''(v)], \label{eq:Ufunction}\\
    \sigma &= \sigma[\phi(v),\phi'(v),\, \psi(v),\psi'(v)]. \label{eq:Sigmafunction}
\end{align}

Since we focus on axisymmetric configurations, the domain needs to be discretized only along the $v$-direction into $N$ nodes,  
$v=\{v_1,v_2,\dots,v_N\}$, with a uniform step size $s=(v_N-v_1)/(N-1)$.  
Denoting $\phi(v_i)\equiv \phi_i$ and $\psi(v_i)\equiv \psi_i$, the first and second derivatives are approximated using finite differences:
\begin{align}
    F'_i &= \frac{F_{i+1}-F_i}{s}, \label{eq:F'}\\
    F''_i &= \frac{F_{i+1}+F_{i-1}-2F_i}{s^2}, \label{eq:F''}
\end{align}
with $F=\phi,\psi$ and $i=2,3,\dots,N-1$.  
For boundary nodes, we set $F'_N = F'_{N-1}$, $F''_1 = F''_2$, and $F''_N = F''_{N-1}$, which provides sufficient accuracy for small step sizes $s$.

Substituting Eqs.~\eqref{eq:F'} and~\eqref{eq:F''} into Eq.~\eqref{eq:Ufunction}, we obtain the discrete total strain energy,
\begin{equation}
    \mathcal{E}_{\Sigma} = \sum_i \mathcal{E}[\phi_{i-1},\phi_i,\phi_{i+1},\, \psi_{i-1},\psi_i,\psi_{i+1}].
\end{equation}

Given the domain size $v_0$, sheet thickness $t$, and the elastic parameters,  
we assign an initial guess $\Phi_i\!\to\!\phi_i$, $\Psi_i\!\to\!\psi_i$, and minimize $\mathcal{E}_{\Sigma}$ with respect to $\phi_i$ and $\psi_i$:
\begin{equation}
    \min \mathcal{E}_{\Sigma} \;\Rightarrow\; \phi_i,\, \psi_i.
\end{equation}
The corresponding numerical stresses follow as
\begin{equation}
    \sigma = \sigma[\phi_i,\phi_{i+1},\, \psi_i,\psi_{i+1}].
\end{equation}

\begin{figure*}[t]
\centering
\includegraphics[width=1.0\linewidth]{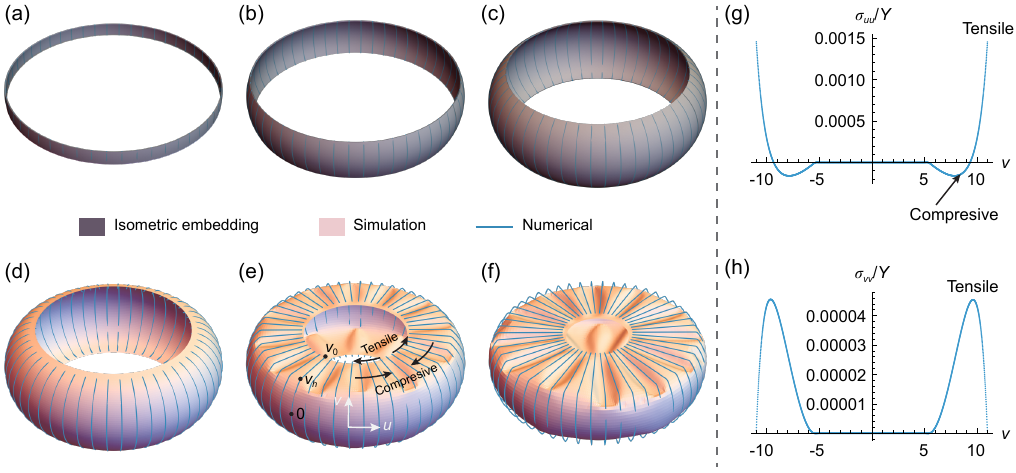}
\caption{\label{figS:FigS6} Comparison among isometric embedding, finite-element simulation, and numerical solution. (a–c) For growth within the isometric regime ($\bar{K}_T < 4\pi$), the results of the analytical isometric embedding, finite-element simulation, and numerical solution overlap, indicating fully compatible configurations. (d-f) Beyond the isometric limit ($\bar{K}_T > 4\pi$), the numerical solutions and finite-element simulations remain consistent at the macroscopic level. Numerical solutions exhibit flattened end regions due to the surface-of-revolution assumptions. (g-h) Normalized in-plane stress components $\sigma/Y$ along the azimuthal ($u$) and meridional ($v$) directions for the configuration in (e). Along the latitude ($u$-direction), stresses near the edge are tensile, transitioning to compressive farther from the rim, whereas along the meridian ($v$-direction) the stresses are tensile throughout. The interior isometric region remains nearly stress-free. These stress distributions account for the flattened ends and the emergence of dimple patterns observed in simulations.  }
\end{figure*}

For $v_0 < v_h$, we use the analytical isometric embedding solutions~\eqref{eq:isometricSol1}–\eqref{eq:isometricSol2} as initial guess.  
For $v_0 > v_h$, the minimization is performed incrementally:  
we divide the "extra" domain size into $M$ steps, $\Delta = (v_0 - v_h)/M$, and use the converged configuration at size $v_h + (j-1)\Delta$ as the initial condition for the next step $v_h + j\Delta$, with $j = 1,2,\dots,M$.  
This continuation approach ensures numerical convergence and captures the evolution of post-limit configurations (see Fig.~\ref{figS:FigS6}).

\section{Topological Surgery}\label{sec:SM-TS}

This analysis elucidates how changing the topology from cylindrical to planar removes the global isometric frustration, thereby restoring the possibility of smooth isometric embedding.

In this section, we perform a topological transformation between cylindrical and planar manifolds to illustrate the transition from frustration to isometric embedding.  
Specifically, we consider a planar manifold 
$\mathcal{M}_p = [0,2\pi]\times[-v_0,v_0]$, 
which can be obtained by cutting a cylindrical surface $\mathcal{M}$ along the $v$-direction.  
Assigning the same reference first and second fundamental forms as in Eq.~\eqref{eq:abar-bbar}, we postulate the following parametrization for an isometric embedding:
\begin{equation}
    \mathbf{X}_p(u,v) = 
    \big(\phi_p(v)\cos(\eta u),\, \phi_p(v)\sin(\eta u),\, \psi_p(v)\big),
\end{equation}
where $\eta$ is a dimensionless parameter that characterizes the azimuthal overlap in the $u$-direction.  
Here, $\eta=1$ corresponds to a non-overlapping configuration, while $\eta>1$ represents an overlapping embedding.

The corresponding first and second fundamental forms read
\begin{equation}
    a =
\begin{pmatrix}
\eta^2\phi_p^2 & 0\\
0 & \phi_p'^2+\psi_p'^2
\end{pmatrix}, \label{eq:ap}
\end{equation}
\begin{equation}
    b =
\begin{pmatrix}
-\dfrac{\eta^2\phi_p\psi_p'}{\sqrt{\phi_p'^2+\psi_p'^2}} & 0\\[6pt]
0 & \dfrac{\phi_p''\psi_p'-\phi_p'\psi_p''}{\sqrt{\phi_p'^2+\psi_p'^2}}
\end{pmatrix}. \label{eq:bp}
\end{equation}

For an isometric embedding, setting $a=\bar{a}$ yields
\begin{align}
    \phi_p(v)  &= \frac{A}{\eta\sqrt{K_0}}\cos(\sqrt{K_0}v), \label{eq:phip}\\
    \psi_p(v)  &= \frac{1}{\sqrt{K_0}}E\!\left(\sqrt{K_0}v\,\Big|\,\frac{A^2}{\eta^2}\right). \label{eq:psip}
\end{align}

Substituting Eqs.~\eqref{eq:phip} and~\eqref{eq:psip} into the energy functional~\eqref{eq:NonEuElasticityEnergy}, we obtain the total bending energy,
\begin{equation}
    \mathcal{E}_b(\eta) = \frac{t^3}{3}
    \int_{\mathcal{M}_p}\|b-\bar{b}\|^2\,dS.
    \label{eq:Ub}
\end{equation}

Minimizing $\mathcal{E}_b(\eta)$ with respect to $\eta$, i.e., $\frac{\partial \mathcal{E}_b(\eta)}{\partial \eta} = 0$, yields
\begin{equation}
    \eta = A.
\end{equation}

Hence, we find an isometric embedding of the same reference manifold but with planar topology:
\begin{equation}
\label{eq:Xp}
    \mathbf{X}_p(u,v) = 
    \big(\phi_p(v)\cos(A u),\, \phi_p(v)\sin(A u),\, \psi_p(v)\big),
\end{equation}
where
\begin{align}
    \phi_p(v) &= \frac{1}{\sqrt{K_0}}\cos(\sqrt{K_0}v), \label{eq:phip1}\\
    \psi_p(v) &= \frac{1}{\sqrt{K_0}}\sin(\sqrt{K_0}v). \label{eq:psip1}
\end{align}

\begin{figure*}[t]
\centering
\includegraphics[width=0.8\linewidth]{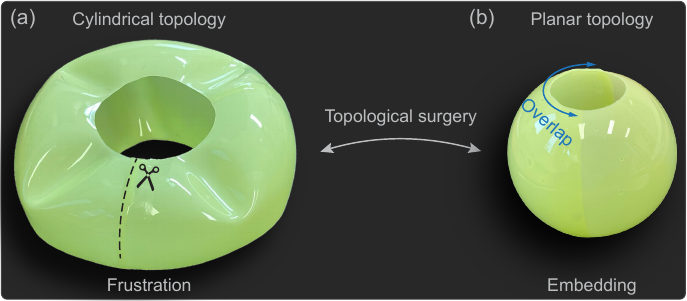}
\caption{\label{figS:FigS7} Topological surgery between cylindrical and planar topologies.  
(a) A frustrated configuration with cylindrical topology. The surface exhibits persistent dimples arising from the topological frustration, which cannot be removed by any local perturbation within the same topology.  (b) An isometric embedding obtained after performing a topological surgery: cutting the bulge surface along a meridian to transform its topology from cylindrical ($\chi=0$) to planar ($\chi=1$).  This topology-relaxing operation releases the global frustration and allows a smooth embedding, albeit with overlapping boundaries. Conversely, gluing the two edges of the planar configuration along the meridian restores the original cylindrical topology and reinstates the frustration.  The parameters used in (a,b) correspond to Eqs.~\eqref{eq:Kspherezone}--\eqref{eq:KTspherezone} with $R_0 = 20$, $r_0 = 10$, $K_0 = R_0^{-1}$, and $A = 1.5$.}
\end{figure*}

\begin{figure*}[t]
\centering
\includegraphics[width=0.8\linewidth]{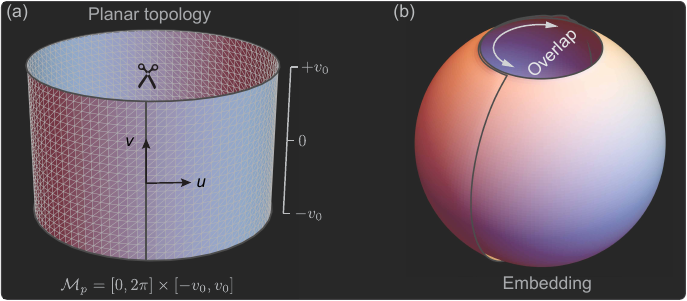}
\caption{\label{figS:FigS8} Isometric embedding releases frustration through topological surgery in simulations. (a) Relaxing the 2D manifold from a cylindrical topology ($\chi=0$) to a planar topology ($\chi=1$) by cutting along the $v$-direction. The domain size $v_0$ and the reference geometry are identical to those in Fig.~\ref{figS:FigS5}c. (b) An isometric embedding is obtained by growing the surface along $v$-direction. As the surface grows, it gradually folds into an open spherical patch with an overlap factor $\eta=A$ along the $u$ direction. All geometric and mechanical parameters remain the same as in the frustrated cylindrical configuration, yet the change of topology removes the global isometric frustration and restores a smooth isometric embedding, corresponding to Eq.~\eqref{eq:Xp}.}
\end{figure*}

This represents an open spherical patch of radius $K_0^{-1/2}$, 
which overlaps by a factor $A$ along the $u$-direction.  
Such a configuration can be realized both experimentally—by cutting the bulged surface along the meridian ($v$-direction) to release the frustration—and numerically by relaxing the system from cylindrical to planar topology (see Figs.~\ref{figS:FigS7} and~\ref{figS:FigS8}). This topological surgery thus bridges the frustrated cylindrical embedding and the strain-free planar isometric embedding. More importantly, it demonstrates the topological nature of the isometric incompatibility identified in this study.

\section{Disc Topology}

\begin{figure*}[t]
\centering
\includegraphics[width=1.0\linewidth]{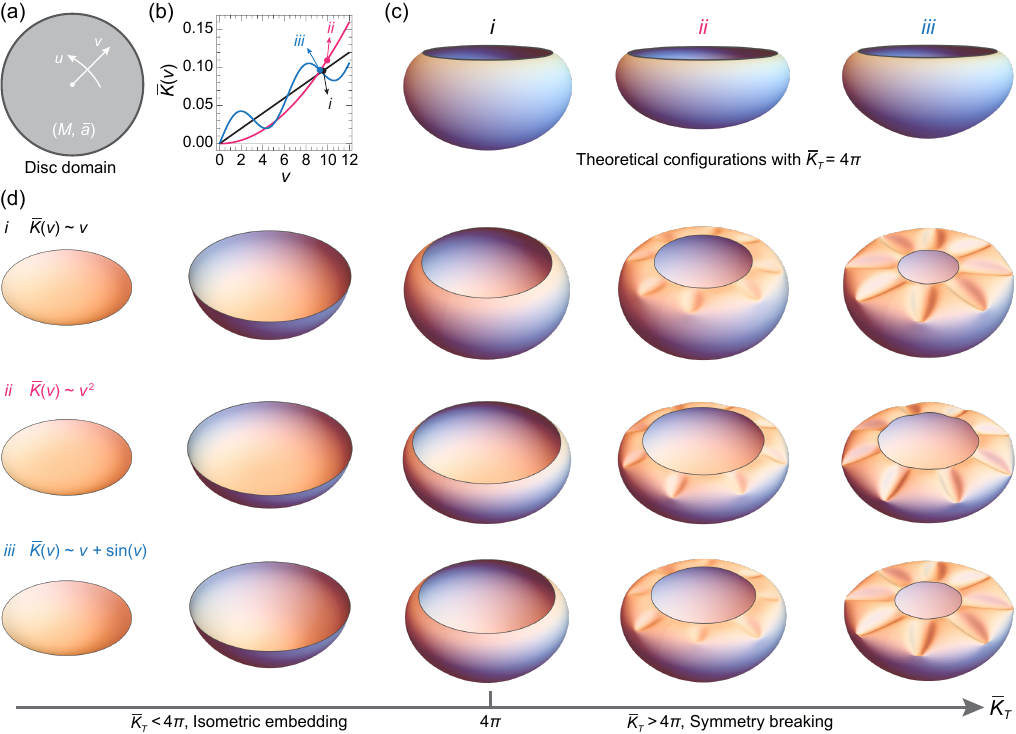}
\caption{\label{figS:FigS9} Isometric embedding and isometry-incompatibility in a disc domain. (a) A disc domain $\mathcal{M}$ endowed with a reference metric $\bar a$ that varies along the radial coordinate $v$. (b) Three representative reference Gaussian curvature profiles $\bar K(v)$ associated with $\bar a$ (Eqs.~\eqref{eq:discK1}-\eqref{eq:discKSin}): ($i$) $\bar{K}(v) = K_0\frac{v}{v_l}$, ($ii$) $\bar{K}(v) = K_0(\frac{v}{v_l})^2$, and ($iii$) $\bar{K}(v) = K_0(\frac{v}{v_l} + \alpha \sin({\beta \frac{v}{v_l}})$. Here, $K_0 = {1}/{100}$, $\alpha = 5/2$, $\beta = 1$ and $v_l = 1, 3$ and $1$, respectively. (c) Axisymmetric theoretical configurations at the isometric-embedding limit $\bar K_T=4\pi$ for the different curvature profiles shown in (b). (d) Simulated growth sequences for the disc domain following the reference metrics in (a)--(b). In all cases, once the reference total Gaussian curvature exceeds $4\pi$, the smooth axisymmetric branch terminates and no $W^{2,2}$ isometric embedding has been identified~\cite{lewicka2011scaling}, independent of the detailed form of $\bar K(v)$.}
\end{figure*}

In this section, we consider a disc domain $\mathcal{M} = [0,2\pi)\times[0,v_0]$ endowed with a reference metric $\bar{a}$ that prescribes a spatially varying Gaussian curvature (Fig.~\ref{figS:FigS9}). According to Eqs.~\eqref{eq:refMetric} and~\eqref{eq:refK}, the reference Gaussian curvature field $\bar{K}(v)$ is given by
\begin{equation}
    \bar{K}(v) = - \frac{\Phi''}{\Phi}.
\end{equation} 

Restricting to surfaces of revolution and imposing an isometric embedding condition, i.e.\ $a=\bar{a}$ and $K(v)=\bar{K}(v)$, and following Eq.~\eqref{eq:X}-\eqref{eq:generalEQ2} leads to

\begin{align}
&\phi'' + \bar{K}(v)\,\phi = 0, \label{eq:discIso1}\\
&\psi(v) = \int_0^{v}\sqrt{1-\phi'^2(v)}\,dv. \label{eq:discIso2}
\end{align}

Similar to the cylindrical topology (cf. Eqs.~\eqref{eq:p'/psi'} and \eqref{eq:RefKT}), 
for axisymmetric isometric embeddings of a disc topology we obtain

\begin{equation}
\label{eq:discp'/psi'}
    \frac{d p}{d \psi} = 2\pi \frac{d \phi}{d \psi} 
    = 2\pi \frac{\phi'(v)}{\sqrt{1-\phi'^2(v)}},
\end{equation}

and

\begin{equation}
\label{eq:discRefKT}
    \bar{K}_T
    = \int_{\mathcal{M}} \bar{K}(v)\sqrt{\det\bar{a}}\;dudv
    = 2\pi\big(\phi'(0)-\phi'(v_0)\big).
\end{equation}

Without loss of generality, we consider the following representative forms of the reference Gaussian curvature $\bar{K}(v)$

\begin{align}
&\bar{K}(v) = K_0 \frac{v}{v_l}, \label{eq:discK1}\\
&\bar{K}(v) = K_0 \left(\frac{v}{v_l}\right)^2, \label{eq:discK2}\\
&\bar{K}(v) = K_0\!\left(\frac{v}{v_l} + \alpha \sin\!\left(\beta\frac{v}{v_l}\right)\right). \label{eq:discKSin}
\end{align}

Combining Eqs.~\eqref{eq:discIso1}–\eqref{eq:discIso2} with 
Eqs.~\eqref{eq:discK1}–\eqref{eq:discKSin}, and imposing the boundary conditions 
$\phi(0)=0$ and $\phi'(0)=1$, the resulting configurations $\mathbf{X}(u,v)$ can be obtained 
analytically and/or numerically. Assuming that the embedding terminates at 
$v_0=v_h$, we have $\phi'(v_h)=-1$. Eqs.~\eqref{eq:discIso2}–\eqref{eq:discRefKT} 
then imply that for $v_0<v_h$, where $\phi'^2(v)<1$ and $\bar{K}_T<4\pi$, smooth 
surfaces of revolution exist. As $v_0\to v_h$, where $\phi'(v_h)\to -1$ and 
$\bar{K}_T\to 4\pi$, a geometric horizon emerges with $\psi'(v_h)\to 0$ and 
$\frac{dp}{d\psi}\big|_{v_h}
=2\pi\frac{d\phi}{d\psi}\big|_{v_h}\to\infty$.

Focusing on the thin-sheet limit, we set $\bar{b}=0$ in the simulations for simplicity; this approximation has negligible influence on the frustrated configurations, as verified in Fig.~\ref{figS:FigS5}. The geometric parameters are chosen as $K_0=1/100$, $\alpha=5/2$, $\beta=1$, and $v_l=1,3,$ and $1$, respectively, while the mechanical parameters are $Y=1000$, $\nu=0.3$, and $t=0.02$. The resulting configurations are shown in Fig.~\ref{figS:FigS9}. These results demonstrate that symmetry breaking occurs once the rigidifying curve forms at $\bar{K}_T=4\pi$, independent of the specific functional form of the reference curvature field $\bar{K}(v)$.

\section{Perspective from the Gauss--Bonnet Theorem}

For both cylindrical (annular) and disc topologies undergoing axisymmetric growth, the Gauss--Bonnet theorem provides a natural geometric interpretation of the emergence of the geometric horizon and the associated $4\pi$ threshold,

\begin{equation}
\int_{\mathcal{M}} \bar{K}\, d\bar{S} + \int_{\partial\mathcal{M}} \bar{k}_g\, d\bar{l}
= 2\pi \chi(\mathcal{M}),
\end{equation}

where $\chi(\mathcal{M})$ is the Euler characteristic of the domain $\mathcal{M}$ ($\chi=1$ for a disc and $\chi=0$ for a cylindrical/annular topology). As the domain grows (e.g., increasing $v_0$), the total reference Gaussian curvature 
$\bar{K}_T = \int_{\mathcal{M}} \bar{K}\, d\bar{S}$ increases. For a given topology this increase should be compensated by the integrated boundary geodesic curvature
$\bar{K}_{\partial} = \int_{\partial\mathcal{M}} \bar{k}_g\, d\bar{l}$, where 
$\bar{k}_g = \frac{\partial_v\bar{a}_{uu}}{2\bar{a}_{uu}\sqrt{\bar{a}_{vv}}}\big|_{v_0}$. 
This relation provides a global geometric interpretation of the horizon condition derived in the main text.

\textit{Disc topology}. As $\bar{K}_T \rightarrow 4\pi$, the Gauss--Bonnet theorem implies $\bar{K}_{\partial} \rightarrow -2\pi$, corresponding to the limiting value attainable by a planar closed curve. Under axisymmetric growth the boundary remains circular, indicating that the edge has reached a geometric horizon. Since $\bar{K}_{\partial}$ cannot decrease below $-2\pi$ during the axisymmetric growth, the total curvature is bounded by $\bar{K}_T \le 4\pi$. At this limit $\phi'(v_h)\to -1$, leading to singular behavior of the embedding at the boundary, including $dp/d\psi\big|_{v_h}\to\infty$ and $b_{vv}\big|_{v_h}\to\infty$. The boundary therefore becomes a rigidifying curve that prevents further continuation of the axisymmetric isometric embedding.

\textit{Cylindrical/annular topology}. Under symmetric growth the two boundaries reach the horizon simultaneously, each contributing $\bar{K}_{\partial\pm}=-2\pi$, yielding the threshold $\bar{K}_T=4\pi$. Further growth induces divergences of $dp/d\psi$ and $b_{vv}$ on the rigidifying boundary $v=v_h$, ruling out axisymmetric isometric embeddings.

Importantly, the value $\bar{K}_T=4\pi$ represents an upper bound rather than a unique threshold. For the asymmetric growth of a cylindrical domain, divergences and symmetry breaking may occur already at $\bar{K}_T<4\pi$. For example, if the $+v$ edge reaches the horizon first ($\bar{K}_{\partial+}=-2\pi$), further growth remains possible in the $-v$ direction due to the remaining geodesic-curvature budget ($\bar{K}_{\partial-}>-2\pi$), while growth in the $+v$ direction is obstructed by the rigidifying curve ($b_{vv}\to\infty$, $dp/d\psi\to\infty$). Symmetry breaking therefore first occurs near that edge.

\begin{figure*}[t]
\centering
\includegraphics[width=1.0\linewidth]{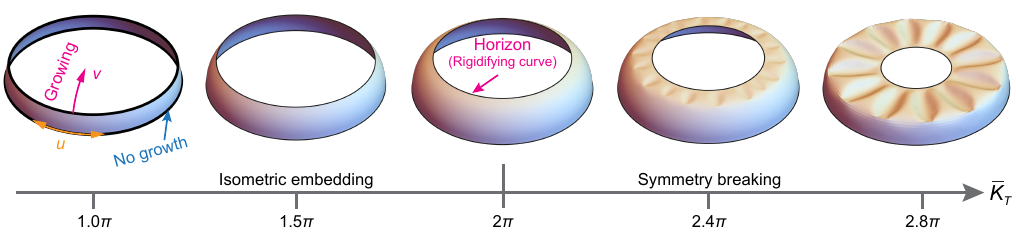}
\caption{\label{figS:FigS10} 
Asymmetric growth of a cylindrical domain producing a hemisphere-like surface. The domain grows from the equator along the $+v$ direction while the lower edge remains stagnant. At $\bar{K}_T = 2\pi$, the upper edge reaches a geometric horizon and becomes a rigidifying curve. Further growth leads to symmetry breaking localized near the upper edge, while the region near the lower edge remains smooth.}
\end{figure*}

To illustrate this mechanism, we consider an asymmetric growth of a cylindrical domain that produces a hemisphere-like surface. Starting from the equator, the domain grows only along the $+v$ direction, while the lower edge remains stagnant (Fig.~\ref{figS:FigS10}). As $\bar{K}_T \to 2\pi$, the upper edge reaches a geometric horizon and forms a rigidifying curve, marking the maximal extent of smooth axisymmetric isometric embedding in the current condition. Further growth leads to symmetry breaking localized near the upper edge, while the region near the lower (non-rigidifying) edge remains smooth. These results demonstrate that the onset of the horizon and symmetry breaking can occur below the global upper bound $\bar{K}_T=4\pi$, and are instead governed by the local exhaustion of the available geodesic-curvature budget on the boundaries ($\bar{K}_{\partial+} = -2\pi$ at the upper edge, while $\bar{K}_{\partial-} = 0$ at the equator). All simulations are performed under free boundary conditions: The reference domain $\mathcal{M} = [0,2\pi)\times [0,v_0]$ encoded with the reference geometry shown in Eq.~\eqref{eq:abar-bbar}; The parameters are set as $A=1.5$, $K_0 =0.01$, $t=0.02$, $Y = 1000$ and $\nu = 0.3$.

\section{Relation to Liebmann’s theorem.}

Liebmann’s theorem states that the only complete, connected surfaces of constant positive Gaussian curvature that admit a smooth isometric immersion in $\mathbb{R}^3$ are round spheres~\cite{do2016differential}. This classical result provides an obstruction for extending positively curved surfaces into complete, boundary-less geometries.

However, the setting considered here differs from the classical setting addressed by Liebmann’s theorem: we address incomplete surfaces with boundaries, generated by growth or prescribed metric on finite domains (discs or annuli). In this case, we show that beyond a critical total Gaussian curvature, isometric embeddings seem to lose $W^{2,2}$ regularity due to the emergence of a rigidifying curve (horizon) along the growing edge.

We emphasize that Liebmann’s theorem remains fully consistent with our results: while it rules out complete smooth embeddings of positively curved metrics beyond the sphere, our findings suggest that the corresponding obstruction already manifests itself at the level of finite open sheets. In this sense, the geometric horizon and the associated loss of regularity may be viewed as a finite-domain precursor of the global obstruction described by Liebmann’s theorem.


\bibliography{SM}